# Shubnikov de Haas quantum oscillation of the surface states in the metallic Bismuth Telluride sheets

Taishi Chen, Junhao Han, Zhaoguo Li, Fengqi Song*, Bo Zhao, Xuefeng Wang, Baigeng Wang, Jianguo Wan, Min Han, Rong Zhang, Guanghou Wang

National Lab of Solid State Microstructures, Nanjing University, Nanjing, 210093, China

## Abstract

Metallic $Bi_2Te_3$ crystalline sheets with the room-temperature resistivity of above 10 mΩ cm were prepared and their magnetoresistive transport was measured in a field of up to 9 Tesla. The Shubnikov de Haas oscillations were identified from the secondly-derived magnetoresistance curves. While changing the angle between the field and normal axis of the sheets, we find that the oscillation periods present a cosine dependence on the angle. This indicates a two-dimensional transport due to the surface state. The work reveals a resolvable surface contribution to the overall conduction even in a metallic topological insulator.

* e-mail songfengqi@nju.edu.cn, Phone: +86 25 83686745, Fax: +86 25 83595535

## 1 Introduction

Three-dimensional topological insulators (TIs) has been acquired great attention in recent years because of its novel electronic structure ,which features a band insulator in its bulk and a metallic gapless Surface State (SS) [1-6]. Such an electronic configuration has been verified by mapping the electronic structure using angle-resolved photoemission spectroscopy (ARPES)[7-10]. Spin helicity has also been demonstrated for the Dirac-type SS [11], which suppresses the impurity-induced back-scattering and leads to a dissipationless carrier transport. Based on such topological SS, the Majorana fermion, dissipationless spintronic devices and even quantum spin hall effect are investigated. $Bi_2Te_3$ and its related thermoelectric family are such TIs. Due to the strong spin-orbit interactions, the electronic bands from Bi and Te (or Se) atoms cross and construct a topology-nontrivial junction near the Γ point in the k space [1]. The interfacial evolution between a TI material and a normal insulator (e.g. vacuum) requires closing the band gap, therefore the upper conduction band and the bottom valance band kiss through an SS in the bulk bandgap [12]. This essentially results in a Dirac-fermion dominated SS as predicted. Intense efforts are currently made to achieve the quantum device based on the topological SS.

Identifying the signature of the SS out of the bulk electronic transport is one of the crucial tasks towards the device applications of TIs since the electronic transport is the vital work before the practical device application [13-21]. One can clearly see the whole band dispersion from the ARPES measurements including SS and Bulk State (BS), while it is still hard to benefit from the SS transport and fabricate the topological devices because of very small SS contributions to the overall conductions in most experimental cases. Firstly, the inevitable Te/Se vacancies offer a P/N type doping to the materials of $Bi_2Te_3$ families, leading a conducting bulk channel [22]. Secondly, upon the exposure of the air and light, the Fermi level will move even deeper into the bulk conduction band [23, 24]. Furthermore, despite the expected highly mobile and free-of-backscattering transport of the SS electrons, the penetration depth of the SS is normally less than a few nanometers [24, 25]. The ratios of the SS conduction are therefore as low as 0.3 percent in the bulk $Bi_2Te_3$ materials, unless changing the composition. It remains less than 10 percent even if we reduce the thickness to less than 100nm. Very complex chemical vapor co-deposition of Bi, Se, Te and Sb has to be carried out in a recent experiment with a SS-dominated transport [26]. Such co-deposition will normally induce a lower electronic mobility [26-28]. The angle-dependent Shubnikov de Haas (SDH) oscillation was proposed as the manifestation of the two-dimensional (2D) SS [18, 19, 29]. Upon a strong magnetic field (B), Landau levels are formed within the Dirac cone. The change of B manipulates the occupation of the landau levels and leads to the SDH oscillation of the overall conduction. It is periodic over 1/B. The unique character of the SS's SDH oscillation is its 2D behavior, where the 1/B period of the SDH oscillation solely depends on the normal component of the TI sheets. Most data on the topological SS conductions are extracted from the SDH analysis since then.

However, till now, the 2D SDH oscillations are mostly observed in bulk-insulating samples [18, 29, 30]. The SDH oscillations can also be found in some well-crystalline metallic samples, but they mostly exhibit three-dimensional behavior. It is proposed that the 2D transport from the topological SS is screened out and the surface carriers might be scattered by the bulk carriers in the metallic samples [18, 30]. Here, we report such evidence of resolvable SS transport in metallic TI samples, where the metallic $Bi_2Te_3$ crystals are prepared and the SS transport is identified by the two-dimensional SDH oscillations. [31, 32]

## 2 Experimental

Single crystals of $Bi_2Te_3$ were prepared by melting high-purity Bi and Tellurium powders (5N, all from Alfa Aesar) and annealing in sealed quartz ampoules. Five batches of crystals were prepared with varying Bi / Te ratios and heating conditions to optimize the carrier concentrations and mobility. The samples were then obtained by mechanical exfoliation with a blade, which dimensions are normally tens of millimeters. The thickness of the samples is as small as a few micrometers. Such flakes are crystalline sheets. As shown in the inset of Fig 1(c), six-probe contacts were made by depositing the room temperature-cured silver paste on the 1mm×1mm regions. The measurements of electronic transport between 10 and 300 K were then conducted using a commercial cryostat with a Keithley 4200 semiconductor analyzer. A typical measurement current is 0.1mA. The magnetoresistance (MR) measurements at 5K were performed in a Quantum Design Physical Property Measurement System. The Hall response was measured at the field of 1T.

## 3 Results

In Figure 1(a), we show a typical temperature profile of the furnace temperature during the preparation procedure. The temperature was rapidly increased to 1000°C and stirred for a few hours. Keeping the stirring, the temperature was decreased to 500°C to allow the crystalline nucleation in 5 days. A 5 day's annealing from 500°C to 420°C finalized the preparation before dropping down to room temperature. The X-ray powder diffraction demonstrates its genuine crystal-line condition of no. 166 space group as referred to the PDF card 820358. The resultant crystals can be easily cleaved to the sheets with the dimensions of a centimeter as shown in Fig. 1(b). We carried out transport measurements to give the resistivity and Hall constants in our lab. A typical room-temperature resistivity distribution in a crystalline crump is shown in Fig 1(b).

We can see a few sheets near the wall of the glass ample are of some low resistivity of a few mΩ·cm, while the highest resistivity of 32.3 mΩ cm (Sample-I) appears in the center of the crump. The samples selected in this study are located between them with a medium resistivity of around 10 mΩcm (Sample-M).The typical temperature dependence of the longitudinal electrical resistivity and carrier density is shown in Fig.1(c). Sample-I is of negative temperature dependence. This shows that we can obtain the samples with an open bandgap (Sample-I) for the study of ideal 3DTIs. Its low-temperature resistivity reaches 0.1Ωcm, which is a satisfactory value in current TI research. Considering the reported values of the mobility and carrier density of the SS, one might expect a good SS contribution in such samples [27]. The resistivity of the Sample-M presents positive temperature dependence with a metallic behavior and falls to around 1mΩcm. The SS contribution therefore becomes minor. We can also see the carrier density of the samples in Fig 1(c), which is around $10^{18}/cm^3$ and similar to the reported values. The mobility of the bulk carriers can also be calculated from the Hall measurement, which is around $2500 cm^2/Vs$. We here successfully resolve the two-dimensional SDH oscillations from the SS in such samples as shown below.

Figure 2(a) shows a typical MR curve of Sample-M. The resistivity increases by 8 times while increasing the field B from 0 to 9T. Interestingly to note, the MR measured with B||c axis displays a nearly linear increase versus B. Such linearity that the previous MRs of metallic $Bi_2Te_3$ did not show has been highlighted as a signature of the anomalous SS transport [18]. It has also been attributed to the competition and compensation between electron and hole bulk bands dominated in $Bi_2Te_3$, forming an interesting topic on its own and being still a matter of debate. No signature of quantum SDH oscillations can be obviously seen. However, when we check the secondly differentiated MR curve in Fig. 2(b), the SDH oscillations appears. A tiny oscillation is seen in the low-field range. Such an oscillation seems to exhibit a period of 0.25T even if we tilt the samples. It is recently related to the Sondheimer Oscillation [31]. Another stronger oscillation rises from 2 T. Its amplitude and beat keep growing while increasing the magnetic field, as seen in the typical SDH oscillations. The Landau levels can be formed at around 2T, by which we expect a mobility of 5000 $cm^2/Vs$ of the conducting electrons.

Figure 3(a) shows the angle-dependent SDH analysis, which demonstrates the 2D behavior of the SDH oscillations. We plot the secondly derived MR values versus 1/B and then find uniform periods. When the field is normal to the sheet (θ=0), the period is $0.054T^{-1}$. It decreases to nearly 0.03 $T^{-1}$ when the sample is tilted by 45° (θ=45°) and continuously decreases during the sample tilting as shown in Fig 3(a). The periodicity can be confirmed by the Fourier-transformation as shown in the inset of Fig 3(b). A dominant period appears in the Fourier-transformed MR curves, in which the dominant frequency increases while θ increases to 45°. The clear 2D signature is demonstrated by multiplying B with cosθ as shown in Fig 3(b). The dominant frequency of the SDH oscillation increases with the angle initially, while all the oscillation frequencies coincide with each other within the measurement uncertainties after we transformed B to the normal component of the magnetic field. This is clear evidence on the 2D transport due to the topological SS, as demonstrated by Qu et al [18]. Reasonably, the contribution from the metallic bulk becomes dominant while θ grows large, over 45°. Please see the SDH oscillations and its FFT curve at 67.5° in the inset of Fig 3(b), which shows two peaks. The weaker one shown by the blue arrows is attributed to the SS. The other peak belongs to the bulk, which persists within the measurement uncertainties and is shown by the red dashed lines. We have to note that the amplitudes in either the secondly derived MR values or the FFT curves are not related to the intensity of the conductance contributions. This allows only ratios of amplitudes to be analyzed. The thickness of the sample is a few micrometers, while its length/width is a few millimeters. The SS is predominant while θ is below 45°, while it becomes much smaller than the bulk state (around 18 Tesla in all the FFT curves) at larger θ. If we carefully check the FFT curves at low θ, we will find the tiny feature from the bulk state. Such coexistence of the SDH oscillations from both a 2D SS and the bulk channel is common in the samples of TIs.

**4 Discussions**

Detailed analysis of the transport environment was carried out for the samples to obtain the ratio of SS conduction. For the bulk transport of Sample-M, their resistivity decrease from over 10 mΩ cm to around 1 mΩ cm while the temperature decreases from 300K to 5K. In the meantime, the mobility of the samples increases from 200 $cm^2/Vs$ to around $3000 cm^2/Vs$ as determined by the Hall measurement as shown in Fig 1(c). The dominant carriers are determined to be electrons and its volume density is around $10^{18} cm^{-3}$. The transport parameters of the SS can be extracted since the SDH oscillation exhibits the 2D behavior which is confirmed in the last paragraph. The dominant frequency of 18.5T in Fig 3(b) leads to the areal carrier density of $4.6 x 10^{11} cm^{-2}$. Fitting the SDH oscillations according to the Dingle's analysis [19], the effective mass of $0.09m_e$ and the Dingle's temperature of 3.2±1.0K are obtained. This leads to a mean free path of 45nm, and the SS mobility of $6825 \pm 2100 cm^2/Vs$[34]. The error can be large due to very small amplitude of the SDH oscillations. Longitudinal zero-magnetic conductance ($G_{xx}$ $(B)|_{B=0T}$) ,which is G(0), can be acquired from 2 dimensional Dirac gas transportation theory, which is mentioned in Qu's article. According this theory, G(0) has the following relations with wave vector number ($k_f$ ) and elastic scattering free-path( l ), namely $G(0)= (e^2/h) k_f l$ , which also can be logically thought as surface state conductance $Gs= (e^2/h) k_f l$ if we focuses only on the surface State of Sample-M. With $k_f l$ ~13.5, we get $G_s$= $5.184\times10^{-4}$Ohm$^{-1}$ while $R_{bulk}$=0.17 Ohm at 10K.Thereby, the surface contribution to the overall conduction is then about 0.01 percent in Sample-M.

As a discussion, the SS transport in the present samples falls between those in the fully insulating and highly metallic crystals. We compared our date to that of the Qu's insulating sample, Q2, which mobility is 10600cm$^2$/Vs and the surface areal density is 7.2x10$^{11}$cm$^{-2}$. The ratio of surface conduction is 0.3 percent [18]. Till now, all reported values for surface carrier density and mobility are around 5x10$^{11}$cm$^{-2}$ and over 5000cm$^2$/Vs [14, 18, 35, 36]. If this is the case for the metallic sample (N1) in Qu's work, the ratio of its surface conduction is around 1ppm. Therefore, obvious 2D SDH oscillations can only be seen in the MR curves of insulating samples, while the weaker one possibly may appears for a highly-metallic sample. Reasonably, we resolve the 2D SDH signals by the way of second derivative on the raw MR data in the sample-M. We also note that high mobility of bulk carriers seem to pose a negative impact on the SDH oscillations of the SS. Both N1 samples and the samples of Butch et al present the mobility of over 10000cm$^2$/V [18, 30], while the samples with 2D SDH oscillations have smaller values of mobility [18, 29]. This may suggest that the higher bulk mobility is sure of enhancing the possibility of arriving to the surface and therefore does harms to the SS transportation. In other words, the 2DTISS transportation can survive over the whole conductance if the surface mobility is much higher than the bulk one, which coincides with our Sample-M, indeed.

## 4 Conclusions

We prepared a series of the TI crystals of $Bi_2Te_3$ by the melting-annealing approach. Some insulating samples reveal dominant SS contributions. We carried out the MR measurement and obtained the linear MR curves in some metallic $Bi_2Te_3$ crystalline sheets with the room-temperature resistivity of above 10 mΩ cm. The SDH oscillations were identified from the secondly-derived magnetoresistance curves and were attributed to the SS by field-tilting measurement. The work reveals a resolvable surface contribution to the overall conduction even in a metallic topological insulator.

**Acknowledgements**

We thank the National Key Projects for Basic Research of China (Grant nos: 2013CB922103, 2011CB922103, 2010CB923400, 2009CB930501), the National Natural Science Foundation of China (Grant numbers: 11023002, 11134005, 60825402, 61176088, 11075076, 11274003), the PAPD project and the Fundamental Research Funds for the Central Universities for financially supporting the work. Helpful assistance from Nanofabrication and Characterization Center, Prof. Yuheng Zhang at High Magnetic Field Laboratory CAS and Dr. Xiang Xiong at Physics College of Nanjing University are also acknowledged.

**Figure Legends**

**Figure 1.** a) the change of furnace temperature during the sample preparation. b) As-prepared refined $Bi_2Te_3$ crystal and exfoliated sheets.. c) The temperature dependence of the resistivity and carrier density (from Hall measurement) of a metallic sheet (Sample-M). The resistance of an insulating sheet (Sample-I) is shown for comparison.

**Figure 2**. a) The MR curve and the inset of b) shows the FFT of the low-field part with the Sondheimer oscillation at $0.5T^{-1}$.

**Figure 3**.Two-dimensional SDH oscillations. a) angle-dependent second-derived MR curves of Sample-M (high-field part). b) analysis of the frequency component in the FFT of the MR curve. The dominant frequency changes systematically with increasing angle. It is constant if the magnetic field is projected to the surface normal.

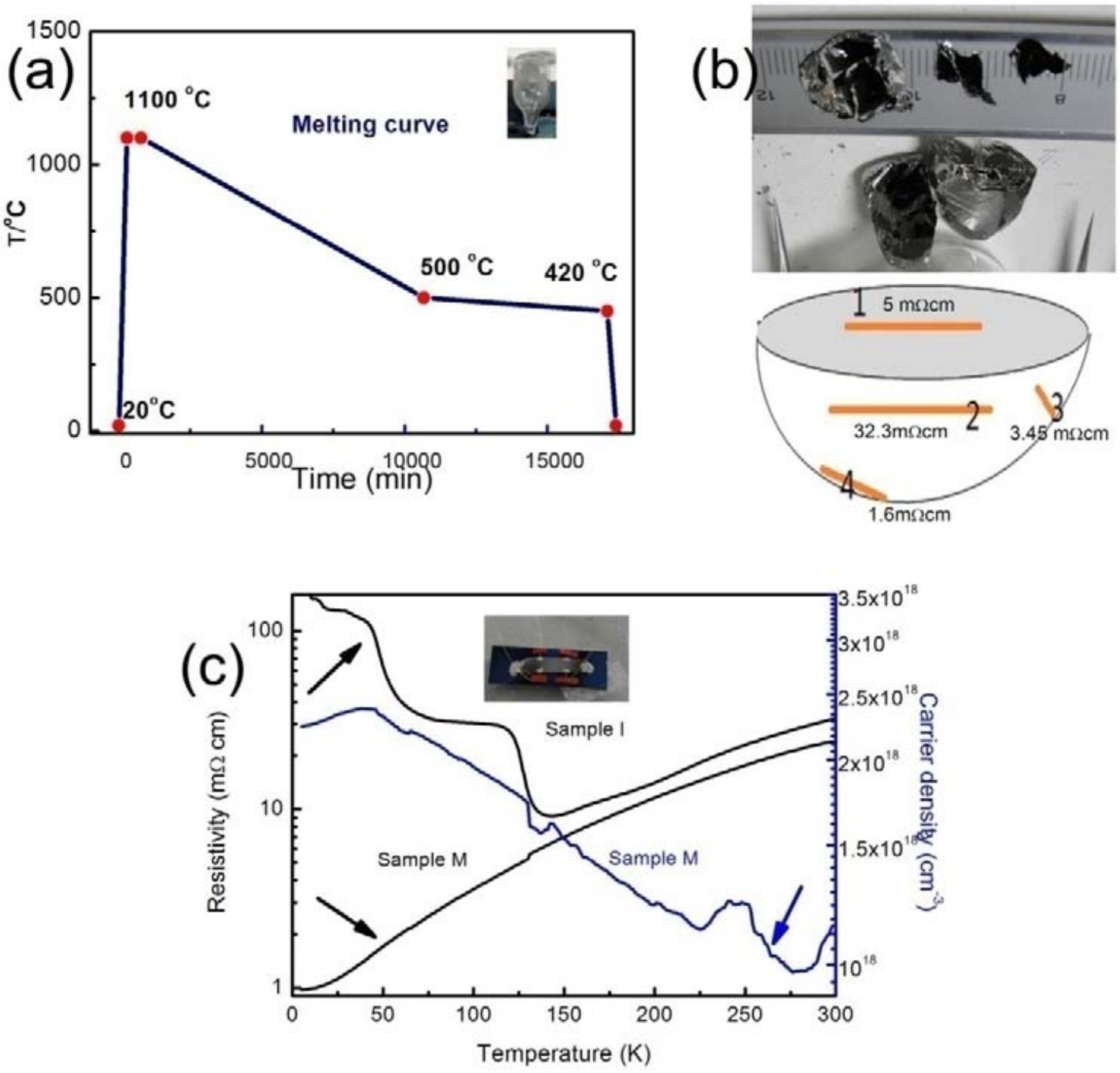

T. Chen et al Figure 1

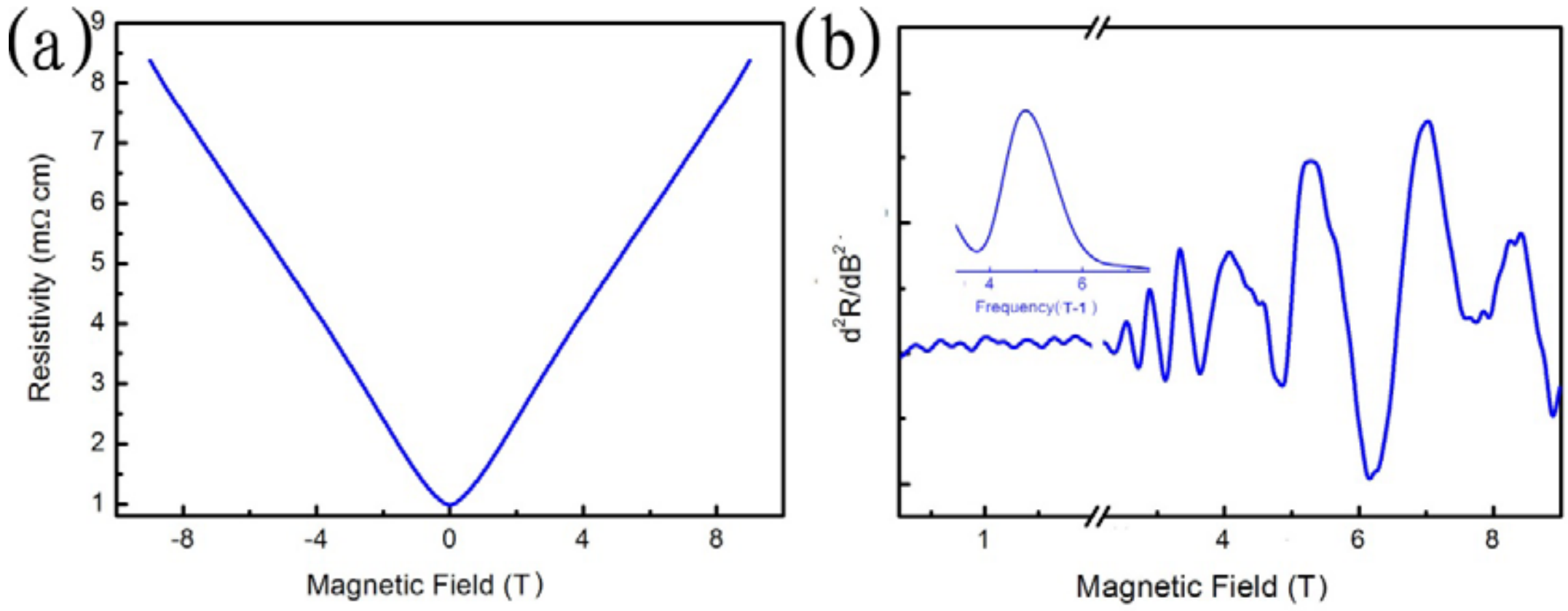

T. Chen et al Figure 2

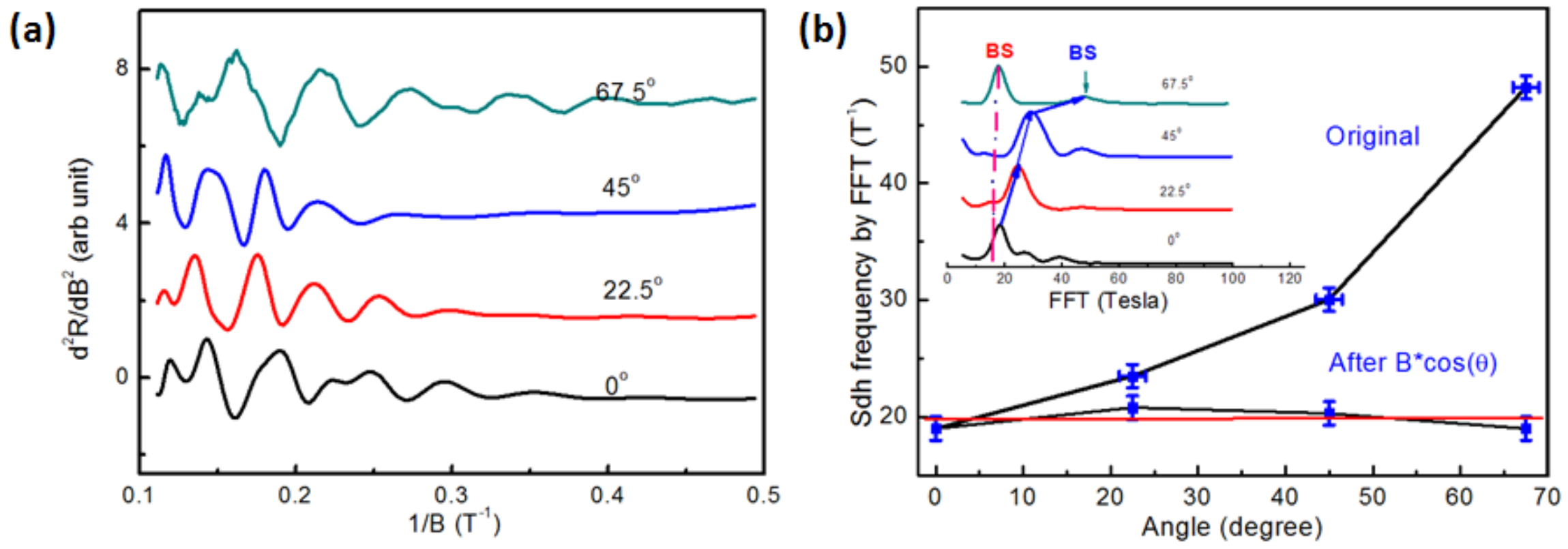

T.

Chen et al Figure 3